\documentclass[aps,showpacs,showkeys,twocolumn,groupedaddress]{revtex4}

\usepackage{amsmath,amsfonts,amssymb,mathrsfs,graphicx,epsfig}

\def\gs{\mathrel{
 \rlap{\raise 0.511ex \hbox{$>$}}{\lower 0.511ex \hbox{$\sim$}}}}
\def\ls{\mathrel{
 \rlap{\raise 0.511ex \hbox{$<$}}{\lower 0.511ex \hbox{$\sim$}}}}

\newcommand{\onbb}{neutrino-less double beta decay}
\newcommand{\ba}{\begin{array}{c}}
\newcommand{\baz}{\begin{array}{cc}}
\newcommand{\bad}{\begin{array}{ccc}}
\newcommand{\bea}{\begin{equation} \begin{array}{c}}
\newcommand{\eea}{ \end{array} \end{equation}}
\newcommand{\ea}{\end{array}}

\newcommand{\dms}{\mbox{$\Delta m^2_{\odot}$}}
\newcommand{\dma}{\mbox{$\Delta m^2_{\rm A}$}}

\newcommand{\be}{\begin{eqnarray}}
\newcommand{\ee}{\end{eqnarray}}

\begin{document}

\vspace*{-1.45cm}
\begin{flushright}
TUM-HEP-626/06; 
hep-ph/0603206
\end{flushright}

\title{
A Supersymmetric Contribution to the Neutrino Mass Matrix \\ and 
Breaking of $\mu$--$\tau$ Symmetry
}
\author{Naoyuki Haba$^{a,b}$}\email{haba@ph.tum.de}
\author{Werner Rodejohann$^b$}\email{werner_rodejohann@ph.tum.de}
\affiliation{ \mbox{ } \\ 
$^a$University of Tokushima, Institute of Theoretical Physics, 
Tokushima-shi 770-8502, Japan \\  \\ $^b$Physik-Department,
Technische Universit{\"a}t M{\"u}nchen, James--Franck--Stra\ss{}e,
85748 Garching bei M{\"u}nchen, Germany}


\begin{abstract}
\noindent
Supersymmetry broken by anomaly mediation suffers from tachyonic  
 slepton masses. 
A possible solution to this problem results in ``decoupling'', i.e., 
the first two generations of sfermions are much heavier 
than the third one. 
We note that in this scenario a sizable loop-induced 
 contribution to the neutrino mass matrix results. 
As an application of this scenario we take advantage of the 
fact that the decoupling evidently not obeys 2--3 generation 
exchange symmetry. 
In the neutrino sector, this 2--3 symmetry (or $\mu$--$\tau$ symmetry) 
is a useful Ansatz to generate zero $\theta_{13}$ and maximal $\theta_{23}$. 
The induced deviations from these values are given for some examples, 
 thereby linking SUSY breaking to the small parameters (including 
possibly the solar mass splitting) of the neutrino sector.

\end{abstract}

\maketitle

\begin{figure}[ht]
\hspace{-2.5cm}
\epsfig{file=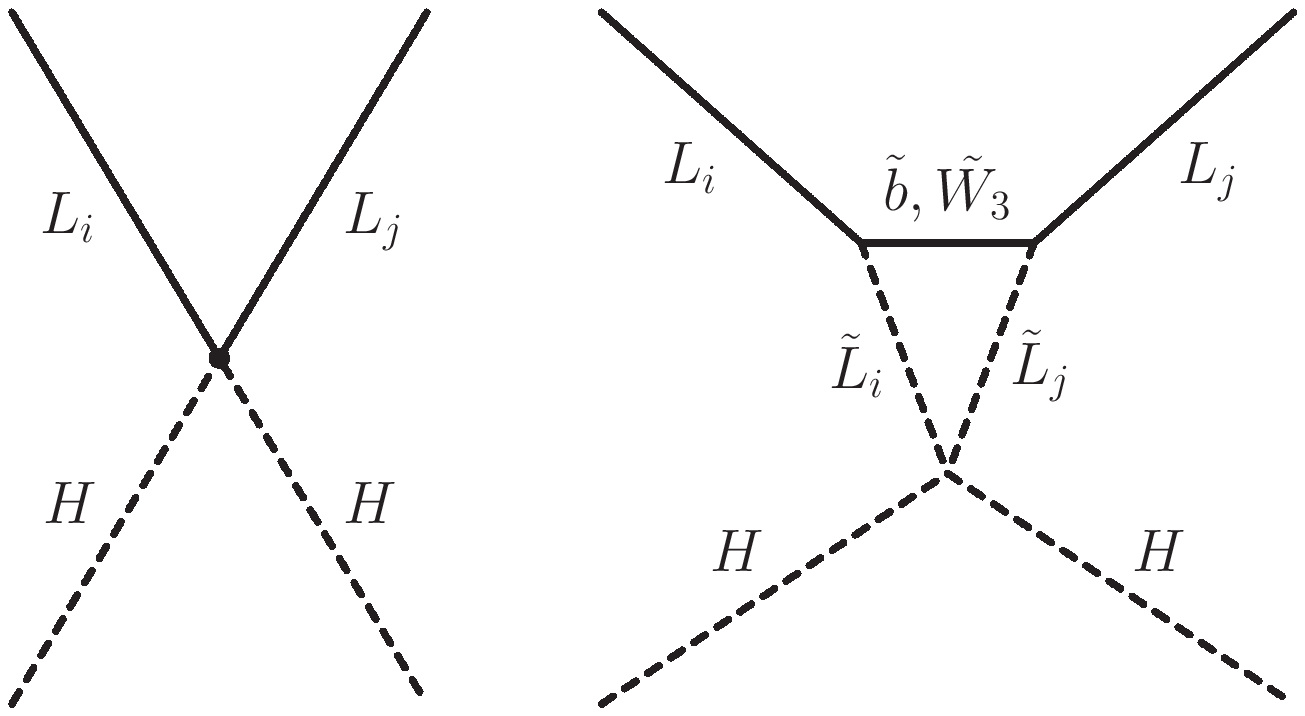,width=11cm,height=11cm}
\vspace{-6.5cm}
\caption{\label{fig:fg}The dimension 5 operator responsible for 
neutrino masses (left, Eq.~(\ref{eq:kappa})) 
and the contribution in the AMSB-decoupling 
scenario (right, Eq.~(\ref{eq:Delta})).}
\end{figure}

Supersymmetry (SUSY) is the best motivated 
candidate for beyond the Standard Model (SM) physics, since it 
for instance provides Dark Matter candidates or solves 
the gauge hierarchy problem. 
Phenomenologically, the requirement of SUSY breaking is obvious, 
its origin and mechanism are however still a mystery. 
One possibility is the anomaly mediated SUSY breaking (AMSB) \cite{AMSB}, 
which always exists in supergravity frameworks. 
It avoids the presence of sizable flavor changing neutral currents, 
the so-called SUSY flavor problem, but has a well-known drawback, namely 
the presence of tachyonic slepton masses. 
One simple solution for this dilemma is obtained by introducing 
an additional $D$-term contribution to the sfermion masses \cite{tree1}. 
Taking into account this contribution can have an interesting effect: 
the 1st and 2nd generation sfermion masses are heavy, 
${\cal O}(1~\rm TeV)$, while the 3rd generation sfermion and 
gaugino masses are ${\cal O}(100~ \rm GeV)$. This latter scale 
corresponds to the scale of the soft SUSY breaking masses 
induced by AMSB, $ \simeq 0.01 \, m_{3/2}$ 
(with $m_{3/2}$ the gravitino mass). 
We denote this framework here the ``AMSB-decoupling'' scenario and will 
place our analysis there. 
In the AMSB scenario the 1-loop correction to 
dimension 5 operators can be of the same order than the tree 
level contribution \cite{Haba:2006dn}. 
One example is the dimension five operator 
responsible for neutrino masses (see Fig.~\ref{fig:fg}): 
\be \label{eq:kappa} 
{\cal L}_{m_\nu}=
\frac{\kappa_{ij}}{2 M_R} \, (\psi_{L_i} \, H)~(\psi_{L_j} \, H)
 +{\rm h.c.}
\ee 
Here $\psi_{L_i}$ is the lepton doublet of flavor $i$, $H$ the up-type 
Higgs doublet and $M_R$ the scale of neutrino mass generation. 
After electroweak symmetry breaking this operator is the neutrino 
mass matrix $m_\nu = \kappa \, \langle H \rangle^2 /M_R$. 
In the present work we will show explicitly how in the 
AMSB-decoupling framework a sizable and flavor-dependent contribution to 
Eq.~(\ref{eq:kappa}) arises and will also discuss some 
interesting applications.

The structure of $\kappa$ defines the neutrino mass and mixing 
phenomena \cite{APSgen} and explaining their peculiar scheme is one 
of the most interesting problems of particle physics. 
An interesting Ansatz is to implement a 2--3 or 
$\mu$--$\tau$ exchange symmetry \cite{mutau} in $\kappa$.  
This yields zero $\theta_{13}$ and maximal $\theta_{23}$, which are 
the best-fit points of global analyzes of neutrino data \cite{thomas}. 
Deviations from these values are achieved by breaking of 
$\mu$--$\tau$ symmetry, which however is usually done 
by hand \cite{mutaubroken,anjan}. 
We propose here to take advantage of the AMSB-decoupling contribution 
to a neutrino mass matrix with $\mu$--$\tau$ symmetry. 
The fact that the first two sfermion families are much heavier than 
the third one apparently violates 2--3 exchange symmetry. 
Assuming that $\kappa$ conserves $\mu$--$\tau$ symmetry and taking the 
$\mu$--$\tau$ violating AMSB correction to $\kappa$ into account 
leads to testable corrections to the initial mixing parameters. 
The size of the 
contribution to Eq.\ (\ref{eq:kappa}) depends on the SUSY masses.
Hence, the implied small 
values of $\theta_{13}$ and $\theta_{23} - \pi/4$ are also sensitive 
to the SUSY masses. In addition, it is also possible to generate the 
small solar neutrino mass squared difference via the AMSB-decoupling 
term.\\

Now let us estimate the 1-loop correction for
 the neutrino dimension 5 operator in Eq.\ (\ref{eq:kappa}). 
As an effect of quantum corrections, the neutrino mass term is obtained   
 by replacing every entry $ij$ as follows: 
\begin{eqnarray} 
(m_\nu)_{ij} \rightarrow (m_\nu)_{ij} \, (1+\Delta_{ij}) \mbox{ for all }
i,j~.
\end{eqnarray}
Here the factor $\Delta_{ij}$ is the correction from the 
 (neutral) wino and bino dressed 
1-loop diagram (see Fig.~\ref{fig:fg}), 
\begin{eqnarray} \label{eq:Delta}
\Delta_{ij} &\simeq& 
 \frac{1}{2} \left( \frac{\alpha_2}{4 \pi} \right)
 F_\phi \, M_2^* \, f^{(2)}_{ij} +
 \frac{3}{10} \left( \frac{\alpha_1}{4 \pi} \right)
 F_\phi \, M_1^* \, f^{(1)}_{ij} \nonumber \\
& \simeq  &
\frac{|M_2|^2}{2} \, (f^{(1)}_{ij}+f^{(2)}_{ij})~.
\label{2}
\end{eqnarray}
We defined the loop-function 
$f^{(I)}_{ij} = f(M_I, \tilde{m}_{{L}_i}, \tilde{m}_{{L}_j})$
as 
\begin{eqnarray}
f(M_I,a,b)\hspace{-1mm}=\hspace{-1mm}
{1 \over a^2-b^2}\hspace{-1mm}
\left[
{a^2 \over a^2-|M_I|^2} \, 
\hspace{-1mm}\ln{a^2 \over |M_I|^2}- (a \leftrightarrow b)
\right]\hspace{-1mm}, 
\end{eqnarray}
where $\alpha_{1,2}$ are the gauge couplings, 
 $M_{1,2}$ the wino ($\tilde{W_3}$) and bino ($\tilde b$) masses, 
 and $\tilde{m}_{{L}_i}$ denotes the mass 
of the $i$th generation slepton. 
$F_\phi$ is the $F$-component of the compensating 
 multiplet of the AMSB sector and is 
 of the order of the gravitino mass.  
The second equation in Eq.~(\ref{2}) uses the 
 AMSB relation  $M_I=(\alpha_I/4\pi)\, b_I \, F_\phi$ 
 ($I = 1,2$ and $b_1=33/5$ and $b_2 = 1$ are the beta functions), 
 which implies $M_2/M_1 \simeq 0.3$. 
In the decoupling mass spectrum under consideration, 
the 1-3 and 2-3 generation sfermion mixings are small, 
so that we can safely neglect their mixings at the vertices. 
Note that we did not specify the origin of the dimension 5 operator 
$\kappa$. Our analysis is therefore independent of the leading 
mechanism of neutrino mass generation. 

The contribution to $\Delta_{ij}$ is of order 
$\alpha_I/(4\pi) \, M_I \, F_\phi \, f^I_{ij}$, 
with $f^I_{ij} \sim M^{-2}$, where $M$ is here some soft 
SUSY breaking mass.  
In other SUSY breaking scenarios there are also diagrams as the one 
displayed in Fig.~\ref{fig:fg}. However, such scenarios typically 
predict that $F_\phi$, the gravitino, gaugino and sfermion 
masses are of the same order of magnitude. 
As a consequence, the loop-diagram gives a 
contribution to $\kappa$ strongly suppressed by a loop factor 
of $10^{-2}$. 
The crucial point we wish to make here occurs when there is {\it only} 
the AMSB breaking effect in which case $F_\phi$ is larger 
by a factor of $10^{2}$ than the gaugino masses. 
Then this factor $10^2$ cancels with $\alpha_I/(4\pi)$ and the 
contribution of $\Delta_{ij}$ to $\kappa_{ij}$ is sizable.

Let us show two typical examples for the magnitude of 
 $\Delta_{ij}$. 
When we take $M_2 = 220$ GeV, 
 $\tilde{m}_{L_3} = 300$ GeV,
 and $\tilde{m}_{L_{1,2}} \geq 4.5$ TeV (``small slepton masses''), 
 the dimensionless matrix $\Delta_{ij}$ 
 is estimated as 
\be 
\label{eq:Deltamnu0}
\Delta_{ij} \simeq 
\left( 
\bad
\lambda^4 & \lambda^4 &  \lambda^3\\[0.2cm]
\cdot &  \lambda^4& \lambda^3 \\[0.2cm]
\cdot & \cdot & \lambda
\ea
\right) \simeq \left( 
\bad
0 & 0 & 0 \\[0.2cm]
\cdot & 0 & 0 \\[0.2cm]
\cdot & \cdot & \lambda 
\ea
\right)~. 
\ee
We introduced here the small parameter 
$\lambda = 0.22$, which is of the order of 
the Cabibbo angle, and proves very useful in the estimates. 
The heavier the 1st and 2nd generation slepton masses become, 
the smaller the elements except for the $(3,3)$ entry become. 
It is the peculiar mass ordering of the sfermions that makes the 
$(3,3)$ entry dominate in $\Delta_{ij}$. 

A similar hierarchy in $\Delta_{ij}$, but with a smaller $(3,3)$ 
entry is obtained by choosing $M_2 =  200$ GeV, 
$\tilde{m}_{L_3} = 760$ GeV, and $\tilde{m}_{L_{1,2}} \geq 8.8$ TeV 
(``large slepton masses''), 
for which 
\be 
\label{eq:Deltamnu2}
\Delta_{ij} \simeq  
\left( 
\bad
\lambda^5 & \lambda^5 &  \lambda^4\\[0.2cm]
\cdot &  \lambda^5 & \lambda^4 \\[0.2cm]
\cdot & \cdot & \lambda^2
\ea
\right) \simeq \left( 
\bad
0 & 0 & 0 \\[0.2cm]
\cdot & 0 & 0 \\[0.2cm]
\cdot & \cdot & \lambda^2
\ea
\right)~. 
\ee 
We can therefore adjust the magnitude of the 
 correction to the neutrino mass matrix by choosing 
 the slepton and gaugino mass spectrum. The sizable differences of 
 the slepton masses and the light gauginos (as typical for AMSB 
scenarios) might be demonstrated at future 
 colliders such as LHC or the ILC.
 
As obvious from Eqs.~(\ref{eq:Deltamnu0}, \ref{eq:Deltamnu2}), the 
correction to $m_\nu$ violates $\mu$--$\tau$ symmetry and can serve 
as a perturbation to such mixing scenarios. 
We will illustrate this in what follows. 
A general $\mu$--$\tau$ symmetric neutrino mass matrix 
leaves the neutrino mass spectrum and the solar neutrino mixing 
angle unconstrained. 
To be definite, we assume here first a normal 
hierarchy and bimaximal mixing \cite{bimax}, for which 
$\theta_{12} = \pi/4$. The mass matrix (with the 
smallest neutrino mass $m_1 \simeq 0$) can be 
written as 
\be \label{eq:mnuNH1}
m_\nu \simeq \frac{m_3}{2}
\left( 
\bad
\epsilon & \epsilon/\sqrt{2} & \epsilon/\sqrt{2} \\[0.2cm]
\cdot & 1 + \epsilon/2 & -1 + \epsilon/2\\[0.2cm]
\cdot & \cdot & 1 + \epsilon/2
\ea
\right) ~, 
\ee
with $\epsilon \simeq \sqrt{\dms/\dma}$, where the 
square root lies at 3 (1)$\sigma$ 
between 0.15 and 0.25 (0.17 and 0.21), with a best-fit value of 0.19 
\cite{thomas}. 
The effect of the AMSB-decoupling contribution in 
Eq.\ (\ref{eq:Deltamnu0}) is now basically just 
multiplying the $(3,3)$ entry of Eq.~(\ref{eq:mnuNH1}) with 
$(1 + \lambda)$. 
We can then predict the modified neutrino observables:  
\begin{eqnarray} \label{eq:resNH1} 
R \equiv \frac{\dms}{\dma} \simeq 
\frac{4\, \epsilon^2 + \epsilon \, \lambda + \lambda^2 /8}
{4 + 2 \, \lambda} ~~,~~|U_{e3}| \simeq \frac{\epsilon \, \lambda}{8}\\ 
\tan \theta_{23} \simeq 1 - \lambda/2 ~~,~~ \nonumber 
\tan 2 \theta_{12} \simeq \frac{4 \, \epsilon}{\lambda}~,
\end{eqnarray}
which for $\lambda=0$ reproduces bimaximal mixing with 
$\dms = \epsilon^2 \, \dma$. For $\lambda \ls \epsilon \simeq \sqrt{R}$ 
the required large solar neutrino mixing 
(observation indicates $\tan 2 \theta_{12} \simeq 2\sqrt{2}$) 
is easily achieved. 
Note that the deviation from maximal 
atmospheric mixing is large, namely of order $\sqrt{R}$, 
whereas $|U_{e3}|$ is of order $R$. Both quantities depend on the 
SUSY breaking parameters as encoded in $\lambda$.

Another possibility for bimaximal mixing and the normal hierarchy 
is when before breaking one chooses $m_1 = -m_2$, i.e., we start with 
vanishing solar $\Delta m^2$. The mass matrix reads
\be \label{eq:mnuNH2}
m_\nu \simeq \frac{m_3}{2}
\left( 
\bad
0 & \sqrt{2} \, \epsilon &  \sqrt{2} \, \epsilon  \\[0.2cm]
\cdot & 1  & -1 \\[0.2cm]
\cdot & \cdot & 1 
\ea
\right) ~, 
\ee
with $\epsilon = m_2/m_3 \ll 1$. 
Note that here the effective mass as measurable in \onbb~(the (1,1) 
element of $m_\nu$) is much smaller than in the previous example. 
Taking now the perturbation from 
Eq.\ (\ref{eq:Deltamnu0}) into account yields 
\begin{eqnarray} \label{eq:resNH2} \nonumber
R \simeq 
\frac{2 \, \epsilon \, \lambda}{4 + 2 \, \lambda} ~~,~~
|U_{e3}| \simeq \frac{\epsilon \, \lambda}{4} ~~,~~
\tan 2 \theta_{12} \simeq \frac{8 \, \epsilon}{\lambda}
\end{eqnarray}
and an identical results for $\theta_{23}$ as in 
Eq.~(\ref{eq:resNH1}). We stress that in this scenario there is 
a link between the small solar mass splitting and the 
breaking of supersymmetry. 

We can also perturb tri-bimaximal mixing \cite{tri}, which is 
$\mu$--$\tau$ symmetry with $\sin^2 \theta_{12} = \frac 13$. 
This requires a smaller perturbation to the mass matrix, i.e., 
Eq.~(\ref{eq:Deltamnu2}), and therefore larger slepton masses.  
Initial tri-bimaximal mixing with $m_1=0$ corrected by 
Eq.~(\ref{eq:Deltamnu2}) yields 
\be
\sin^2 \theta_{12} \simeq \frac 13 + \frac 19 \, \lambda^2/\epsilon~,
\ee 
very small 
$|U_{e3}| \simeq \epsilon \, \lambda^2/\sqrt{72}$ and 
moreover close-to-maximal $\tan \theta_{23} \simeq 1 - \lambda^2/2$. 
$R$ is of order $\epsilon^2$. All deviations from the 
initial values are much smaller than for initial bimaximal mixing. 
In principle it would be possible that the correction to 
the mass matrix stems from Eq.~(\ref{eq:Deltamnu0}), i.e., is larger. 
To accommodate the observed value of $\sin^2 \theta_{12} \simeq \frac 13$, 
however, the contribution proportional to $\lambda$ should be 
suppressed by fine-tuned values of possible $CP$ phases.\\

Consider now the inverted hierarchy. Since in the inverted hierarchy one 
can not perturb the bimaximal mixing to accommodate the observed large 
but non-maximal solar mixing (unless one accepts  
fine-tuning), we start with free $\theta_{12}$:
\bea \label{eq:mnuIH1} \nonumber 
m_\nu \simeq m
\left( 
\bad
A & B & B \\[0.2cm]
\cdot & D & D  \\[0.2cm]
\cdot & \cdot & D 
\ea
\right) \\[0.3cm]
\mbox{ with } 
A \equiv s_{12}^2 - (1 - \epsilon) \, c_{12}^2 ~~,~
B \equiv \frac{1}{2\sqrt{2}} \, (2 - \epsilon) 
\, \sin 2 \theta_{12}  ~~,~\\[0.3cm]
D \equiv \frac \epsilon4 + \frac 14 (2 - \epsilon) \, \cos 2 \theta_{12} 
\eea
and $\epsilon \equiv (m_2 - m_1)/m$, where $m_2 \simeq m_1 \simeq m 
\simeq \sqrt{\dma}$. 
The two non-zero masses $m_1$ and $m_2$ have opposite $CP$ parities here. 
As in the normal hierarchy with tri-bimaximal mixing, 
we require a small perturbation to the mass matrix: 
if we would add Eq.\ (\ref{eq:Deltamnu0}) to this matrix, then 
the ratio of solar and atmospheric $\Delta m^2$ would be of order 
$\lambda + 2 \, \epsilon$, which is too large 
(this will also hold for the unstable case of equal 
$CP$ parities of the masses $m_1$ and $m_2$). 
Hence, we are lead to use Eq.\ (\ref{eq:Deltamnu2}), corresponding to 
large slepton masses.  
Choosing $\theta_{12}$ such that $\sin^2 \theta_{12} = \frac 13$, 
i.e., again tri-bimaximal mixing \cite{tri}, one finds 
\begin{eqnarray} \label{eq:resIH1} \nonumber
R  
\simeq 
\frac{\lambda^2 + 2 \, \epsilon} 
{1 + \frac 23 \, \lambda^2 + \frac 89 \, \epsilon} ~~,~~
\tan^2 \theta_{23} \simeq 1 + \frac 23 \, \lambda^2 ~,\\ 
|U_{e3}| \simeq \frac{\sqrt{2} \lambda^2}{3} ~~,~~ 
\sin^2 \theta_{12} \simeq \frac 13 - 
\frac 19 \, \lambda^2 + \frac{2}{27} \, \epsilon ~.
\end{eqnarray}
The order of magnitude of $\theta_{13}$ is the same as in the case 
of normal hierarchy and initial bimaximal mixing, 
namely of order $R$. The deviation from maximal 
atmospheric mixing is of order $R$. 
Starting with $\epsilon=0$, i.e., with vanishing $\dms$, we have 
$R \simeq 3 \, |U_{e3}|/\sqrt{2}$ and 
\be
\sin^2 \theta_{12} \simeq \frac 13 - \frac{1}{3\sqrt{2}} \, |U_{e3}| \simeq 
\frac 12 - \frac 16 \, \tan^2 \theta_{23}~.
\ee
Again, a large contribution to the mass matrix corresponding to 
smaller slepton masses is in principle possible but requires fine-tuned 
$CP$ phases.\\ 

Up to now it is obvious that the $\mu$--$\tau$ symmetry was only obeyed by 
the neutrino mass matrix, while the charged lepton 
 mass matrix was real and diagonal. 
We choose now an example in which both the 
neutrinos and the charged leptons conserve $\mu$--$\tau$ symmetry. 
Naively, one would expect that even after breaking the symmetry the 
PMNS matrix would contain only a sizable 12 mixing, since the 
(close-to-)maximal 23 angle cancels in the product of the matrices 
diagonalizing the neutrinos 
and charged leptons, respectively. This can change for 
quasi-degenerate light neutrinos \cite{anjan}. The mass matrices are: 
\bea \nonumber 
m_{\ell, \nu} = 
\left( 
\bad 
A_{\ell, \nu} & B_{\ell, \nu} & B_{\ell, \nu} \\[0.2cm]
\cdot & D_{\ell, \nu} + E_{\ell, \nu} & E_{\ell, \nu} - D_{\ell, \nu} 
\\[0.2cm] \nonumber 
\cdot & \cdot & D_{\ell, \nu} + E_{\ell, \nu} 
\ea
\right)~, \\[0.3cm]  \nonumber 
\mbox{ with } 
A_{\ell, \nu} \equiv m_{e,1} \, (c_{12}^{\ell, \nu})^2 + m_{\mu, 2} 
\, (s_{12}^{\ell, \nu})^2 ~~,~ \\[0.3cm] 
B_{\ell, \nu} \equiv (m_{\mu, 2} - m_{e, 1}) \, c_{12}^{\ell, \nu} 
\, s_{12}^{\ell, \nu} /\sqrt{2} ~~,~\\[0.3cm] 
D_{\ell, \nu} \equiv m_{\tau, 3}/2 ~~,~  E_{\ell, \nu} \equiv  \frac 12 
(m_{\mu, 2} \, (c_{12}^{\ell, \nu})^2 + m_{e, 1} \, (s_{12}^{\ell, \nu})^2) 
~.
\eea
Here $m_{e, \mu, \tau}$ are the charged lepton masses and 
$c_{12}^{\ell, \nu}$ ($s_{12}^{\ell, \nu}$) the cosine (sine) of the 
12 mixing angle which diagonalizes the charged lepton and neutrino 
mass matrix, respectively. 
Note that $m_\ell$ is symmetric. That does not affect our 
analysis, since for non-symmetric $m_\ell$ one would have to choose a 
$\mu$--$\tau$ symmetric $m_\ell \,m_\ell^\dagger$, which would give the 
same 
results. 
Perturbing the neutrino mass matrix with the term of 
Eq.\ (\ref{eq:Deltamnu0}), we have in the basis of 
diagonal charged leptons (choosing for 
simplicity $\theta_{12}^\ell \ll 1$)
\bea \label{eq:mnuqd}
m_\nu  \simeq  
\left( 
\bad 
A_\nu 
& \sqrt{2} \,  B_\nu & 0 \\[0.2cm]
\cdot & 2 \, E_\nu  
&  \frac 12 (D_\nu + E_\nu) \, \lambda \\[0.2cm]
\cdot & \cdot &  2 \, D_\nu 
\ea
\right)~,
\eea
where only the leading terms are shown. Large atmospheric neutrino 
mixing is only possible for $D_\nu \simeq E_\nu$. This corresponds to a 
quasi-degenerate spectrum and is exactly the scenario put forward in 
\cite{anjan} to accommodate $\mu$--$\tau$ symmetry in 
both the charged lepton and the neutrino mass matrix. Focusing on the 
$23$ sector and denoting the common mass scale with 
$m_0$, close-to-maximal $\theta_{23}$ is achieved for 
$\lambda\,(D_\nu + E_\nu) \gg |D_\nu - E_\nu|$, which leads to 
$\cos 2\theta_{23} \simeq 4 \, m_0 \, |D_\nu - E_\nu|/\dma $ 
\footnote{Note that the form of the mass matrix Eq.~(\ref{eq:mnuqd}) 
is generated at a scale of order TeV. Hence, the running of the 
mass and mixing parameters to low scale $m_Z$ is 
rather moderate, in particular for small neutrino masses.}. 
The parameter $\lambda$ is related to the masses: 
\be 
\lambda \simeq \frac{\dma}{2 m_0^2} \simeq 
\left\{ 
\bad  
0.11 & \mbox{ for } & m_0 = 0.1~ \rm eV~, \\[0.2cm]
0.012 & \mbox{ for } & m_0 = 0.3~ \rm eV~.
\ea
\right. 
\ee
Such values of the neutrino masses are 
in agreement even with the tightest cosmological constraints \cite{cosmo}. 
Light neutrino masses around 0.1 eV require therefore a 
perturbation with small 
slepton masses and larger neutrino masses need larger slepton masses. 

Numerically, for $m_0 \simeq 0.1$ eV we have 
$\cos 2\theta_{23} \simeq |D_\nu - E_\nu|/(0.0055 \, {\rm eV})$, 
which shows the required fine-tuning in such a scenario. 
If $\theta_{23} = \pi/4 - 0.1$, then $|D_\nu - E_\nu| \simeq 0.0011$ eV and 
for $\theta_{23} = \pi/4 - 0.01$ one has 
$|D_\nu - E_\nu| \simeq 0.0001$ eV. For a larger 
mass of $m_0 \simeq 0.3$ eV one finds that for 
$\theta_{23} = \pi/4 - 0.1$ one requires $|D_\nu - E_\nu| \simeq 0.0004$ 
eV, 
while for $\theta_{23} = \pi/4 - 0.01$ one has 
$|D_\nu - E_\nu| \simeq 0.00004$ eV. \\

In summary, anomaly mediated supersymmetry breaking with connection to 
decoupling of the sfermions can generate 
a sizable contribution to the neutrino mass matrix. This can serve 
as a natural explanation for a perturbation of neutrino mixing scenarios. 
Noting that the decoupling scenario violates 2--3 exchange symmetry, 
it is natural to take $\mu$--$\tau$ 
symmetry as an example. We showed in various 
examples that interesting and testable \cite{sandhya} correlations 
between the neutrino mixing parameters result. 
We stress again that in the scenarios presented here 
the breaking of supersymmetry is intimately related 
to small $U_{e3}$, $\cos 2 \theta_{23}$ and even to the small ratio of the 
solar and atmospheric $\Delta m^2$.

\vspace{0.5cm}
\begin{center}
{\bf Acknowledgments}
\end{center}
The work of N.H.\ was supported by the Alexander-von-Humboldt-Foundation. 
The work of W.R.\ was supported by the ``Deutsche Forschungsgemeinschaft'' 
in the ``Sonderforschungsbereich 375 f\"ur Astroteilchenphysik'' 
and under project number RO--2516/3--1.
We would like to thank T.~Ota and N.~Okada for
 useful discussions.

\end{document}